\documentclass{elsart}

\usepackage{amssymb}
\usepackage{graphicx,psfig,epsfig}

\begin{document}

\begin{frontmatter}



\title{From di-hadron correlations to parton intrinsic transverse momentum
in proton-proton collisions\thanksref{1}}

\thanks[1]{Supported in part by U.S. DE-FG02-86ER40251, U.S. NSF INT-0435701, 
and Hungarian OTKA grants T043455, T047050, and NK062044.}


\author[a]{George Fai}, \author[b]{P\'eter L\'evai}, and \author[c]{G\'abor Papp}

\address[a]{CNR, Department of Physics, Kent State University, Kent OH-44242, USA}
\address[b]{RMKI, 
            P.O. Box 49, Budapest H-1525, Hungary}
\address[c]{Dept. for Theor. Phys., E{\"o}tv{\"o}s Univ., 
	    P{\'a}zm{\'a}ny P. 1/A, Budapest 1117, Hungary}  

\begin{abstract}
Di-hadron correlations in proton-proton collisions at $\sqrt{s} = 200$ GeV
are interpreted in terms of a fragmentation width and a momentum imbalance. 
A fragmentation width of $580 \pm 50$ GeV/c is 
obtained, and the momentum imbalance gives an `intrinsic' transverse momentum
width of partons in the proton of $2.6 \pm 0.2$ GeV/c.
\end{abstract}

\end{frontmatter}

\vspace*{-0.2cm}
\section{Introduction}
\label{sec_int}
\vspace*{-0.3cm}
Recent high-statistics runs at RHIC allow the study of two-particle correlations. 
For Au+Au collisions, the correlations augment information from 
jet quenching, and are a valuable tool of jet tomography. Questions raised by
di-hadron studies motivated programs of three-body correlation 
measurements to clarify the collective dynamics of nuclear collisions.

To ascertain the properties of the correlation data in heavy-ion
collisions a proton-proton ($pp$) reference is needed, preferably 
at the same energy. Correlation data in $pp$ at $\sqrt{s}=200$ GeV 
have become available recently~\cite{unknown:2006sc}. Our goal here is
to provide a physical picture of this rich data set on near and away 
side correlations. To interpret the data,
we want to stay as close as possible to Ref.~\cite{unknown:2006sc} 
with the ingredients of our calculation. 
We focus on pion correlations.  

\section{Model}
\label{sec_mod}
\vspace*{-0.3cm}
The single-pion inclusive production cross section can be written as
\begin{equation}
\frac{d\sigma_{\pi}}{p_T dp_T} = \int \frac{d\sigma_{j}}{p_T dp_T} D_j(z) dz
=\int \frac{d\sigma_{j}}{{\hat p}_T d{\hat p}_T} D_j(z)\frac{dz}{z^2}
=\int f_j D_j(z)\frac{dz}{z^2} \ ,
\end{equation}
\vspace*{-.3cm}
where $d\sigma_{j}/{\hat p}_T d{\hat p}_T$ refers to the differential jet
cross section in terms of the parton transverse momentum ${\hat p}_T$,
and $z$ is the momentum fraction carried by the observed hadron.
The quantity $f_j$ is a parton (jet) distribution averaged over quarks, 
antiquarks, and gluons, and  $D_j(z)$ is an average fragmentation 
function. Using $f_j ({\hat p}_T) \propto {\hat p}_T^{-n}$ and 
$D_j(z) \propto z^{-\alpha} (1-z)^\beta (1+z)^{-\gamma}$ with parameters 
$n=7.4$, $\alpha=0.32$, $\beta=0.72$, $\gamma=10.65$\cite{unknown:2006sc},
we reproduce the measured pion spectra~\cite{Adler:2003pb} for $p_T > 3$ GeV, 
which is satisfactory for our present analysis.

On this basis, we constructed a simple model to describe two-particle 
correlations within a jet and between back-to-back jets starting from a 
hard $2 \rightarrow 2$ parton-parton collision. For 
two pions produced from the same parton (near side correlation)
the two-particle cross section can be written as
\begin{equation}
d\sigma_{\pi_1 \pi_2} = \int_0^1 dz_1 \int_0^1 dz_2
 \ d\sigma_j \ D_{j1}(z_1) D_{j2}(z_2) \ \Theta (1-z_1-z_2) \ ,
\label{2dsig0}
\end{equation}
\vspace*{-.3cm}
(and similarly for the away-side correlation, but without the $\Theta (1-z_1-z_2)$
function), where the momentum fractions $z_1$ and $z_2$ describe the relation 
between the parton and hadron momenta, $p^*_{T1} = z_1 {\hat p}_T$ and
$p^*_{T2} = z_2 {\hat p}_T$. Moreover, due to fragmentation, the produced hadrons
acquire a random transverse momentum component.
In addition, the momentum imbalance ($K_T$) between the produced partons due to 
intrinsic transverse momentum, gluon radiation, or any other $2 \rightarrow 3$ 
process will generate a more complicated kinematic situation, where, in the case 
of back-to-back jets, ${\hat p}_{T1}$ and ${\hat p}_{T2}$ are already not collinear.

Considering hadron 1 as the trigger hadron and hadron 2 as the associated hadron,
after kinematic transformations,
\vspace*{-.3cm} 
\begin{eqnarray}
\frac{d\sigma_{\pi_t \pi_a}}{d p_{Tt} d{\Delta \phi} d p_{Ta} } 
&=& J^* \frac{d\sigma_{\pi_t \pi_a}}{d {p}^*_{Tt} dz_t d{p}^*_{Ta} } 
= J \cdot f_j D_{jt}(z_t) D_{ja}(z_a) \ \Theta(1-z_t-z_a) \, ,
\label{Xdphi}
\end{eqnarray}
where $\Delta \phi$ is the azimuthal angle difference between the trigger 
and associated hadrons, and $J$ and $J^*$ represent the proper
Jacobi determinants of the variable transformations. 

We have developed a Monte-Carlo based calculation to model Eq.~\ref{Xdphi}.
We use Gaussian distributions and uniformly
distributed random angles for the fragmentation transverse momenta 
and the momentum imbalance $K_T$. The widths of these distributions 
that best fit the data\cite{unknown:2006sc} are extracted.

\vspace*{-0.8cm}
\section{Results}
\label{sec_res}
\vspace*{-0.8cm}

\begin{figure}[h]
\begin{minipage}{68mm}
\vspace*{-0.2cm}
\resizebox{68mm}{75mm}
{\includegraphics{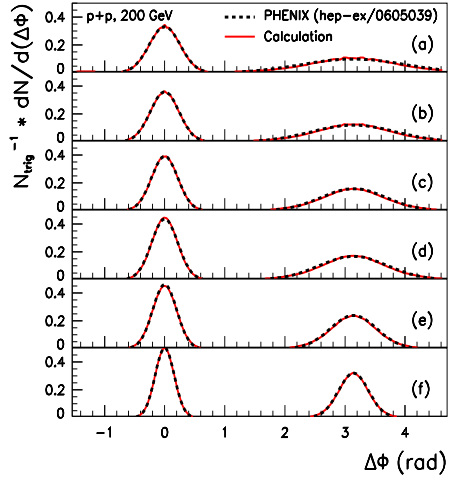}}
\vspace*{-0.8cm}
\caption{\footnotesize
(Color online) Di-hadron correlations in $pp$ collisions at $\sqrt{s} = 200$~GeV 
(full lines) compared to data (dashed)~\cite{unknown:2006sc}. 
}
\label{fig1}
\end{minipage}
\hspace{\fill}
\begin{minipage}{68mm}
We have calculated di-hadron correlation functions in $pp$ collisions at
$\sqrt{s} = 200$ GeV at various trigger and associated transverse momenta
and compared the results to available experimental data\cite{unknown:2006sc} 
to extract the widths of the fragmentation transverse momentum distribution 
and the $K_T$ distribution. A sample comparison is displayed in Fig.~1, where 
the dashed lines indicate a fit through the data points, and the full lines 
represent our calculations. The associated transverse momentum is kept in the 
$1.4 \leq p_{Ta} \leq 5.0$ GeV/c range, 
while the trigger 
transverse momentum is increasing from $2.5 \leq p_{Tt} \leq 3.0$ GeV/c 
to $6.5 \leq p_{Tt} \leq 8.0$ GeV/c moving down through the panels.
The ``pedestal'' of
\end{minipage}
\end{figure}
\vspace*{-0.436cm}
the correlation functions has been cut off.  
The agreement is very good, although small deviations are visible 
in a more magnified view.

Figure 2 shows the values of the widths of the fragmentation and 
`intrinsic' momentum distributions in given $p_{Tt}$ 
windows as functions of $p_{Ta}$. The fragmentation width appears to be 
constant, $\sqrt{\langle j_T^2 \rangle} = 580 \pm 50$ GeV/c, 
independent of $p_{Ta}$ and $p_{Tt}$. The $k_T$ width shows a 
dependence on $p_{Tt}$ (similarly to a simpler treatment~\cite{Levai:2005fa}), 
displayed for two different data sets in Fig.~3. 
 
\begin{figure}[b]
\begin{minipage}[t]{73mm}
\resizebox{73mm}{82mm}
{\includegraphics{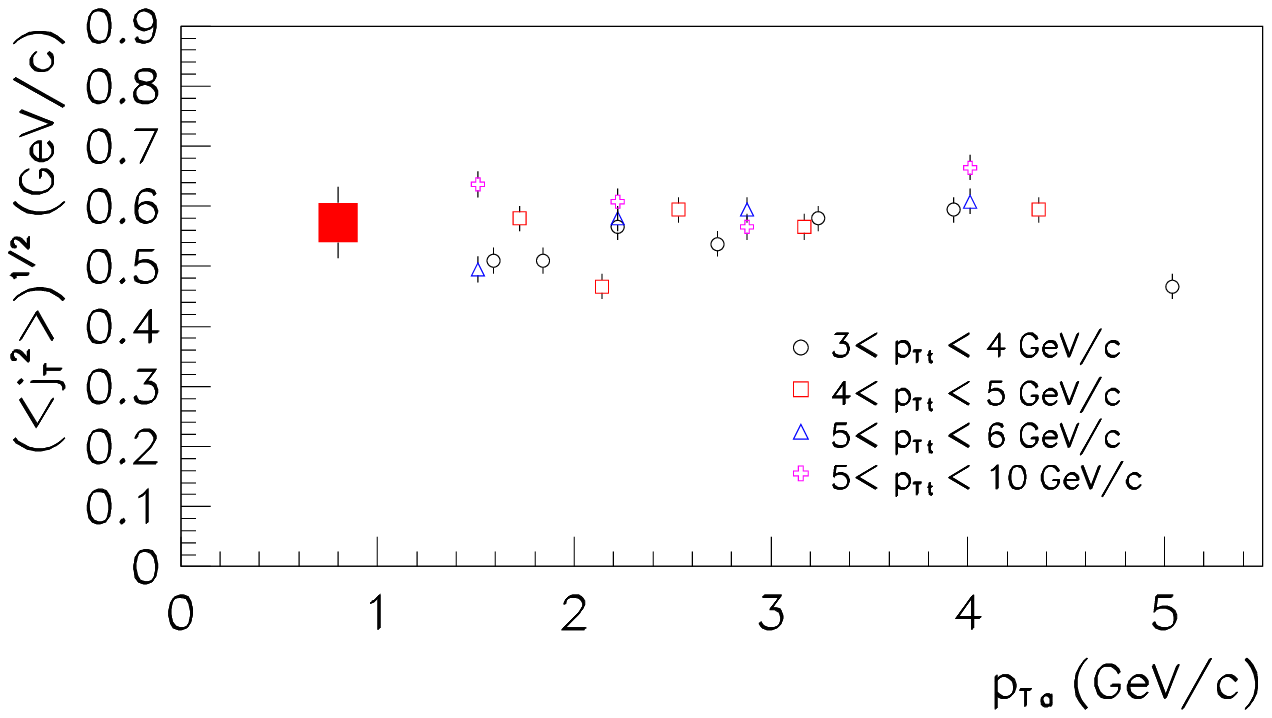}}
\label{fig2}
\end{minipage}
\vspace*{-1.2cm}
\hspace{\fill}
\begin{minipage}[t]{73mm}
\resizebox{73mm}{82mm}
{\includegraphics{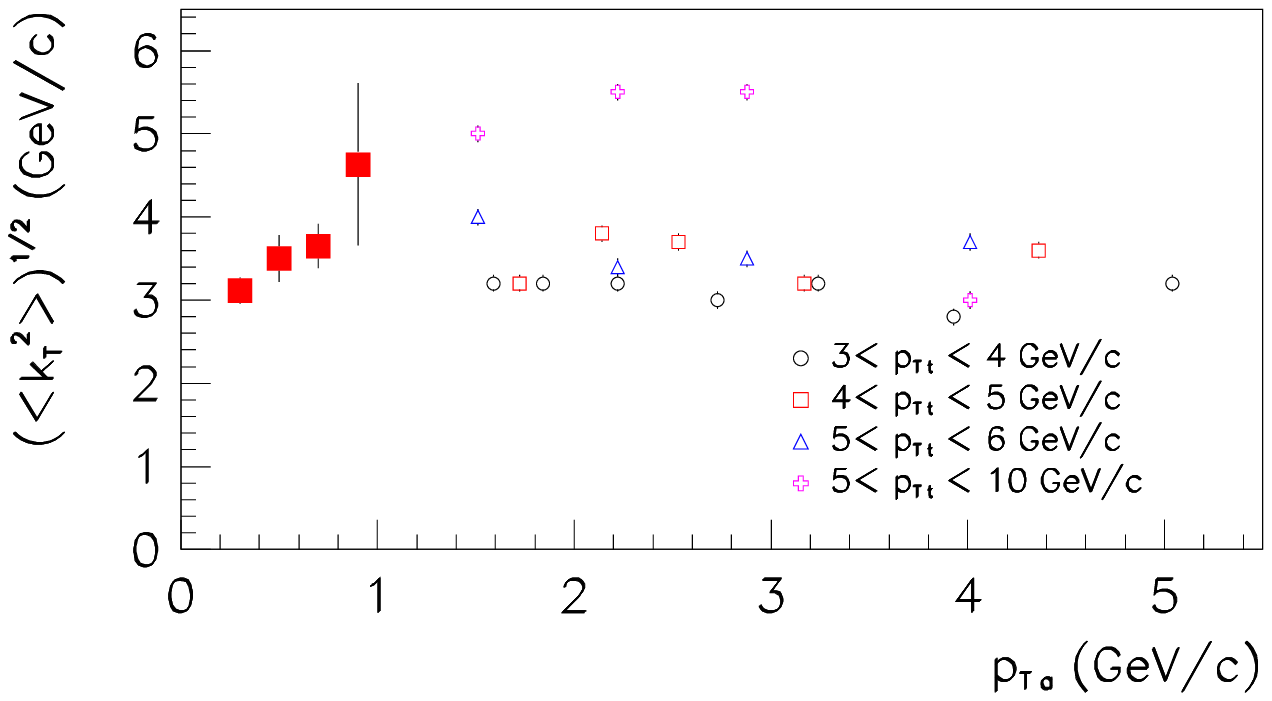}}
\label{fig3}
\end{minipage}
\vspace*{-1.2cm}
\caption{\footnotesize
(Color online) Best fit values of the Gaussian fragmentation width (left)
and $k_T$ width (right) to reproduce the near and away side 
peaks in given $p_{Tt}$ windows as a function of $p_{Ta}$. Averages
for $p_{Tt}$ windows (only the grand total average on the left) are indicated 
by the large filled squares. Data are from Ref.~\cite{unknown:2006sc}.
}
\end{figure}     

\begin{figure}[h]
\begin{minipage}{70mm}
\vspace*{-0.2cm}
\resizebox{70mm}{50mm}
{\includegraphics{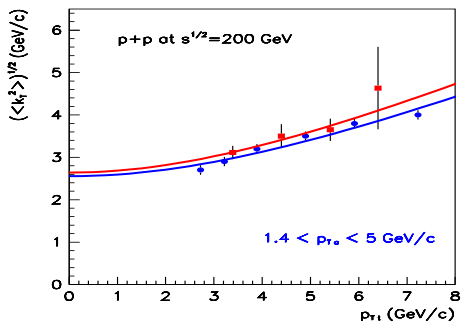}}
\vspace*{-0.5truecm}
\caption{\footnotesize
(Color online) Width of the `intrinsic' transverse 
momentum distribution as a function of the trigger transverse momentum for 
two sets of data.
}
\label{fig4}
\end{minipage}
\hspace{\fill}
\begin{minipage}{65mm}
The spread between the two least-square fitted curves in Fig.~3 is 
indicative of the uncertainty in our procedure.
The behavior of the `intrinsic' transverse momentum width as a function of
$p_{Tt}$ can be understood in terms of its composition\cite{Boer:2003tx}. In addition
to the `true' intrinsic transverse momentum of partons in the proton, there is a 
component from soft gluon radiation that can be handled via resummation, and a 
higher-order contribution which is expected to grow with trigger transverse momentum:
\end{minipage}
\end{figure}
\begin{equation}
\frac{\langle p_T^2 \rangle_{pair}}{2} = \langle k_T^2 \rangle =
\langle k_T^2 \rangle_{intrinsic} + \langle k_T^2 \rangle_{soft} + 
\langle k_T^2 \rangle_{higher-order} \,\, . 
\label{kTcomp}
\end{equation}
\vspace*{-0.2cm}
A measure of the `intrinsic' transverse momentum of partons in the proton (in which
we include the soft gluon radiation component) can be read from Fig.~\ref{fig4}
extrapolating to $p_{Tt} = 0$. This leads to 
$\sqrt{\langle k_T^2 \rangle} = 2.6 \pm 0.2$ GeV/c, in agreement with the value
arrived at in Ref.~\cite{unknown:2006sc}, albeit by a 
different argument. 

\vspace*{-0.5cm}
\section{Conclusion}
\label{sec_conc}
\vspace*{-0.3cm}

We have developed a constructive method for taking into account the fragmentation width
and the momentum imbalance in the treatment of di-hadron correlations. The model has 
the flexibility to treat any $K_T$ distribution. Our results have been tested against
recent PHENIX data on di-hadron correlations in $pp$ collisions. A value of  
$2.6 \pm 0.2$ GeV/c is obtained for the width of the `intrinsic' transverse momentum 
distribution. Future work includes generalization to $pA$ and $AA$ collisions, 
inclusion of the pseudorapidity dimension of the correlations, and application in 
jet quenching calculations.

\end{document}